\documentclass{article}
\usepackage{ijcai19}
\usepackage[english]{babel}

\usepackage{times}
\usepackage{hyperref}
\usepackage{soul}
\usepackage{url}
\usepackage[utf8]{inputenc}
\usepackage[small]{caption}
\usepackage{graphicx}
\usepackage{amsmath}
\usepackage{booktabs}
\usepackage{graphicx}
\usepackage{subfigure}
\title{Reducing Labelled Data Requirement for Pneumonia Segmentation using Image Augmentations}

\author{
Jitesh Seth$^{1,2}$\and
Rohit Lokwani$^{2}$\and
Viraj Kulkarni$^{2}$\and
Aniruddha Pant$^{2}$\and
Amit Kharat$^{2}$
\affiliations
$^{1}$Indian Institute of Science Education and Research, Pune
\affiliations
$^{2}$DeepTek Inc\\
\emails
}

\begin{document}

\maketitle

\begin{abstract}
Deep learning semantic segmentation algorithms can localise abnormalities or opacities from chest radiographs. However, the task of collecting and annotating training data is expensive and requires expertise which remains a bottleneck for algorithm performance. We investigate the effect of image augmentations on reducing the requirement of labelled data in the semantic segmentation of chest X-rays for pneumonia detection. We train fully convolutional network models on subsets of different sizes from the total training data. We apply a different image augmentation while training each model and compare it to the baseline trained on the entire dataset without augmentations. We find that rotate and mixup are the best augmentations amongst rotate, mixup, translate, gamma and horizontal flip, wherein they reduce the labelled data requirement by 70\% while performing comparably to the baseline in terms of AUC and mean IoU in our experiments.

\end{abstract}

\section{Introduction}
The progress in computer vision has had a substantial impact on radiology \cite{McBee2018,Ginneken2017}. Deep learning approaches such as Convolutional Neural Networks (CNNs) have shown great success in the classification and segmentation of radiographs like chest X-rays and CT scans \cite{Lakhani2017,Dunnmon2018}. Research on pathology identification in chest X-rays (CXRs) has mainly focussed on classification, where models predict a class label from a broad set of pathologies. Such an approach does not directly inform us of the regions in the CXRs responsible for the class label. It requires an additional step of plotting saliency maps or gradient class activation maps to interpret how the CNN is making decisions \cite{Pasa2019}.

Segmentation and object detection algorithms provide an advantage over standard CNNs because they can predict the regions of interest. Architectures such as U-Net \cite{Ronneberger2015} and RetinaNet \cite{Lin2018} localise the regions responsible for a specific pathology. Besides, semantic segmentation models are more useful than classification models in two ways: they require less training data since they have pixel-level labels for every image, and they can assist radiologists in their work by localising the abnormalities \cite{Hurt2020}.

However, the cost and effort to label the data is often a significant constraint on training models that perform well in practice \cite{Prevedello2019}. There have been efforts on both fronts - to increase the amount of labelled data available and to create techniques that can learn better from small amounts of labelled data. The former include large databases such as CheXpert \cite{Irvin2019}, and ChestX-ray8 \cite{Wang2017}, whereas the latter consists of image augmentations, semi-supervised learning, special architectures such as U-Nets \cite{Ronneberger2015}, GANs \cite{Goodfellow2014} and others.

Image augmentations is a vital data processing technique to improve the performance of machine learning models. They can make the CNN indifferent to naturally present variations in the data, such as position, scale, or different radiography equipment. Augmentations can be of varied types. Rotate, scale and flip and similar augmentations are called geometrical augmentations. Photometric augmentations transform the colour space of the images. More complex transformations include elastic deformation or mixing multiple images. However, inflating the dataset with numerous augmentations would add to the training time and compute requirements without necessarily increasing performance. A drawback of augmentations is that they may cause overfitting by making the CNN invariant to some features but highly tailored to the training data in others \cite{Shorten2019}.

The problem of augmentations on the performance of segmentation models in the medical domain is not sufficiently addressed in research. Knowledge of specific augmentations which reduce the labelled data requirements will help researchers and data scientists fine-tune their models faster and better. Moreover, data augmentation studies investigate the increase in model performance rather than the decrease in the labelled training data requirements. Our paper specifically addresses the reduction in the training data requirement using image augmentations.

In this paper, we implement five different augmentations on the training of CNN models on Chest X-rays. We propose three criteria for identifying augmentations that reduce labelled data requirement. First, the model should perform comparably to the baseline on a subset of the data. Second, the models with augmentation trained on partial data should perform better than models without any augmentations trained on the same data. Third, the model should satisfy the criteria above for multiple test sets. We validated our results with an in-sample and out-of-sample test set.

\begin{figure*}[htbp]
\includegraphics[width=\textwidth]{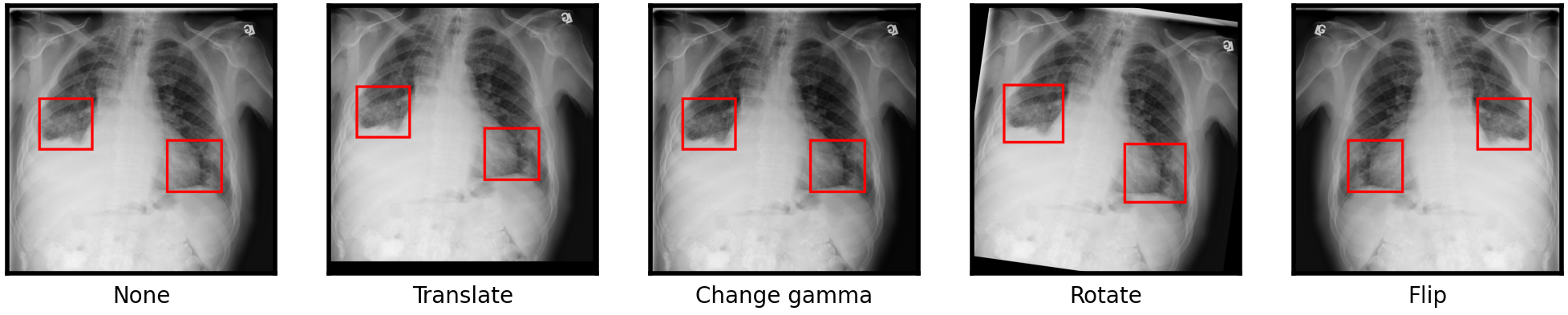}
\caption{Original CXR with opacities labelled(left) and result of augmentations with updated bounding boxes.}\label{aug}
\end{figure*}

\section{Related Work}
This section covers the recent advancements in Computer-aided diagnosis (CADx) using deep learning, image augmentations and semantic segmentation in the field of CXRs.
CADx for chest radiography started around since 1960s \cite{Ginneken2001}, but deep learning has transformed and dominated the field in a few years. Van Ginneken et al. \cite{Ginneken2017} has comprehensively summarised the evolution of CADx from rule-based and machine learning approaches to deep learning ones. Kermany et al. \cite{Kermany2018} showed the generalizability of deep neural networks by using the same neural network to classify retinal Optical Coherence Tomography images and pediatric pneumonia classification in CXRs. The latter model achieved an accuracy of 92.8\%, and the area under the ROC curve was 96.8\%.

Recent studies have proven that image augmentations improve the machine learning model performance. Sirazitdinov et al. \cite{Sirazitdinov2019} showed the effect of augmentations on the classification of chest radiographs. They concluded that a combination of increasing brightness, random rotation and horizontal flips led to the best performance on the ChestX-Ray14 dataset, with an AUC-ROC of 0.808 (compared to 0.785 without any augmentations). However, they do not quantify the extent of each augmentation, thus decreasing its reproducibility. After the invention of mixup \cite{Zhang2018}, Eaton-Rosen et al. \cite{EatonRosen2018} applied it to a dataset of MRI images. They provide a graphical overview of mixup compared to other augmentations and a baseline for a large (199 images) and a small (10 images) dataset.

Souza et al. \cite{Souza_2019} created an automatic method for segmentation and reconstruction of lungs, which can take into account lung opacities from pneumonia or tuberculosis, reconstruct the lung boundaries, and finally, segment the lungs. They used the segmented lungs for a classification model, which achieved an accuracy of 96.97\%, an average Dice coefficient of 0.94 on the Montgomery County’s Tuberculosis Control dataset \cite{Jaeger2014}. Selvan et al. \cite{Selvan2020} tackled the same problem by treating high opacity regions as missing data and using a variational auto-encoder for data imputation. They achieved an accuracy of 88.15\% and a Dice coefficient of 0.8503 on a curated CXR dataset. Thus segmentation was mainly used to demarcate the lungs for use in classification models. However, we can also use semantic segmentation algorithms to demarcate lung opacities. This opportunity became plausible after the publication of the Radiology Society of North America's (RSNA) pneumonia dataset \cite{Shih2019}, also used in this study.

Wu et al. took the concept of lung segmentation and opacity detection one step forward. They segmented both the lungs, divided them into three zones each and predicted the presence of pneumonia in each zone using the patient's radiology report. Thus they created an object detection dataset from radiology reports. Using this dataset, they trained a RetinaNet model and tested it on the RSNA data set. The model had a mean IoU of 0.29 per pneumonia positive image.
Hurt et al. have shown that semantic segmentation of CXRs can be used as a probability map to interpret the radiographs. Their segmentation model on the RSNA dataset showed a dice coefficient of 0.603, and the classification had an AUC of 0.854.

An essential aspect of deep learning in CADx is the usability of the models - we do not desire clever models that might end up being clinically irrelevant \cite{Lundervold2019}. For example, both in Pan and Cadrin-Chênevert's, and Cheng's model of the RSNA dataset \cite{Pan2019}, they systematically decreased the predicted bounding boxes by 12-17\%, which increased the performance for the particular test set, but there is no medically relevant reason to do the same in practical settings. Another practical constraint with Pan's models was extensive ensembling, which requires the availability of high-end GPUs. 

\begin{figure}
    \centering
    \includegraphics[width=0.9\linewidth]{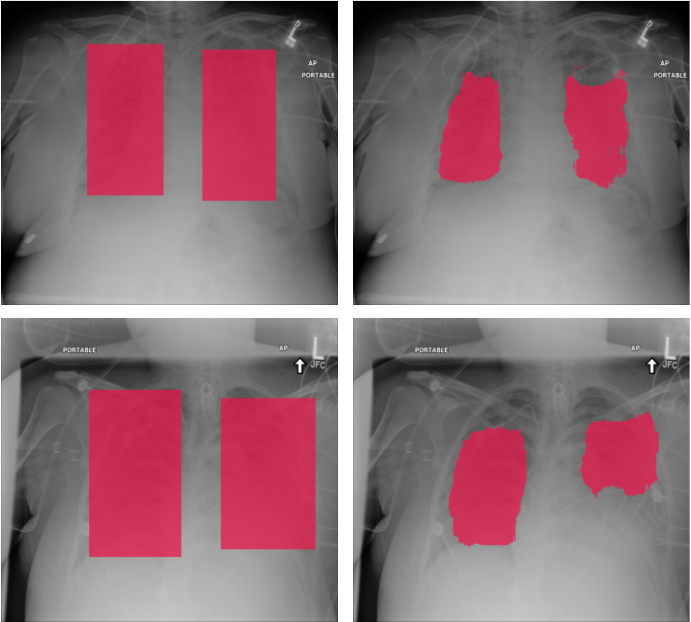}
    \caption{Two examples of CXR with the ground truth mask (left) and predicted mask by the baseline model (right)}
    \label{fig:my_label}
\end{figure}

\section{Data and Methodology}

\subsection{Data}

For this study, we use the publicly available pneumonia dataset jointly annotated by RSNA and the Society of Thoracic Radiology(STR) \cite{Shih2019}. The dataset consists of DICOM images of Chest X-Rays (CXRs) having dimensions $1024\times 1024$. Pneumonia positive images contain bounding box information of lung opacities. The dataset consists of about 30,000 frontal CXRs with bounding boxes around the lung opacities.

From this dataset, we select all the pneumonia-positive CXRs (n=6012) and a subset of the negative CXRs (n=8488) and divide them into training(n=10000), validation (n=1500) and test (n=3000) sets. Each CXR belongs to a unique individual. All three sets have the same prevalence of the positive class (41.4\%).

We also test all trained models on an out-of-sample test set curated from Padchest \cite{Bustos2020} and four private hospitals and population screening programmes from India and Indonesia. This set contains 1125 pneumonia positive CXRs with the corresponding polygonal or rectangular annotations and 1875 pneumonia negative CXRs. 

\subsection{Model Architecture and Evaluation}

We use a U-net-like CNN with depthwise separable convolutions style connections implemented using the Keras library for this study \cite{Team}. We resize the input into an image of shape (512,512,3). The network has two parts, an encoder part and a decoder part. The encoder part uses residual connections, depthwise separable convolution and 2D convolution and max pooling. The decoder portion uses transpose convolution, 2D upsampling and 2D convolution.

We use binary cross-entropy as the loss function, and the evaluation metric is the mean intersection over union (IoU). We use Adam to optimise the loss function and the learning rate is scheduled to decrease when plateauing for 5 epochs. We train each model until it has not shown improvement in validation loss for at least 5 epochs. We save the model weights with the lowest validation loss for evaluation.

After the model training, we evaluate it on the in-sample and out-of-sample test sets. These are compared to the ground truth masks to calculate the mean IoU and the loss. We use the segmentation results to classify the CXR as positive or negative. We calculate the area under the curve (AUC) of the receiver operating characteristics (ROC) based on this classification.

\subsection{Augmentations}

We have chosen five image augmentations for this study. These are random rotation between $-10^{\circ}$ to $10^{\circ}$, changing the gamma between 0.75 to 1.25, translating the image randomly between 0-5\% of its length in x and y direction, horizontal flips, and mixup \cite{Zhang2018}. In mixup, two images and their masks (represented by $x_{1}$ and $x_{2}$) are combined using the formula:

$$x=x_{1}\lambda + x_{2}(1-\lambda)$$

Where $\lambda \sim \beta(0.2, 0.2)$. These augmentations have negligible computational cost. Mixup results in better-performing segmentation models according to recent studies \cite{EatonRosen2018}.

We train the baseline (i.e. no augmentations) on 100\% of the training set. For each of the six conditions - five augmentations and one with no augmentation (hereafter referred to as ``NoAug''), models were trained using 30\%, 50\%, 70\% and 90\% of the training set.

\subsection{Statistical Analysis}
We compare each model with the baseline and the NoAug model on the same amount of data. We use the non-parametric DeLong test \cite{delong1988} to compare the AUC for the models' classification performance. We set the significance level at 0.05. We use the pROC library in R \cite{Robin2011} to perform the same.

\begin{figure}
\centering
\includegraphics[width=0.9\linewidth]{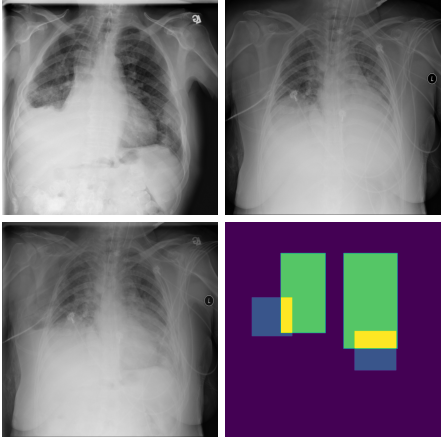}
\caption{(Top) original CXRs for mixup; (bottom left) result of mixup augmentation and (bottom right) corresponding mask.}\label{mixup}
\end{figure}

\begin{figure*}[!htbp]
\begin{center}
\subfigure[]{\label{iou1} \includegraphics[width=0.4\textwidth]{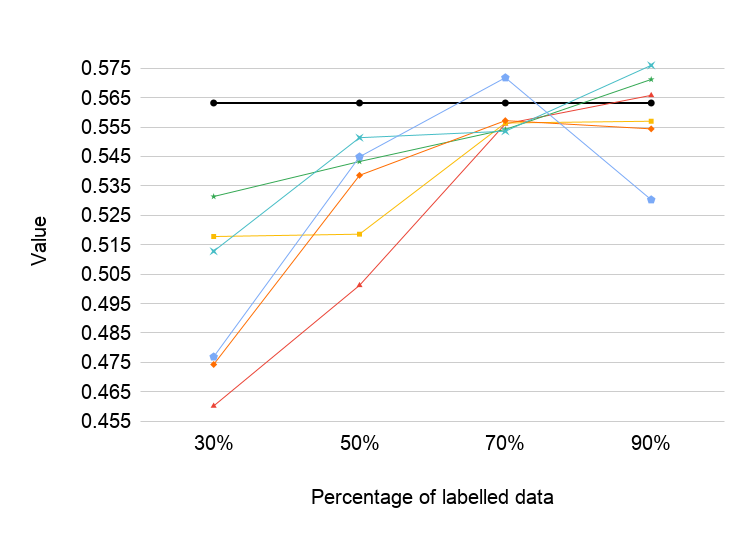}}
\subfigure[]{\label{iou2} \includegraphics[width=0.4\textwidth]{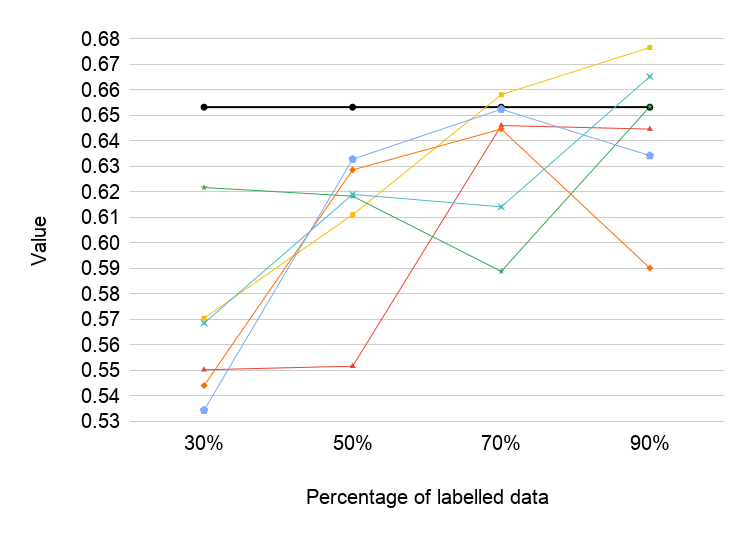}}
\subfigure[]{\label{legend}
\includegraphics[width=0.1\textwidth]{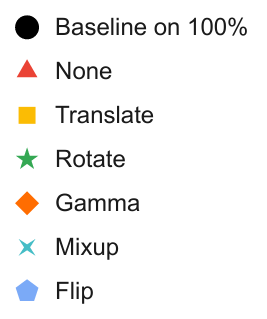}}
\caption{Mean IoU on the \subref{iou1} internal test set and \subref{iou2} external test set for the various models trained. \subref{legend} The legend}\label{iou}
\end{center}
\end{figure*}

\begin{figure*}
\begin{center}
	\subfigure[]{\label{auc1}  \includegraphics[width=0.45\textwidth]{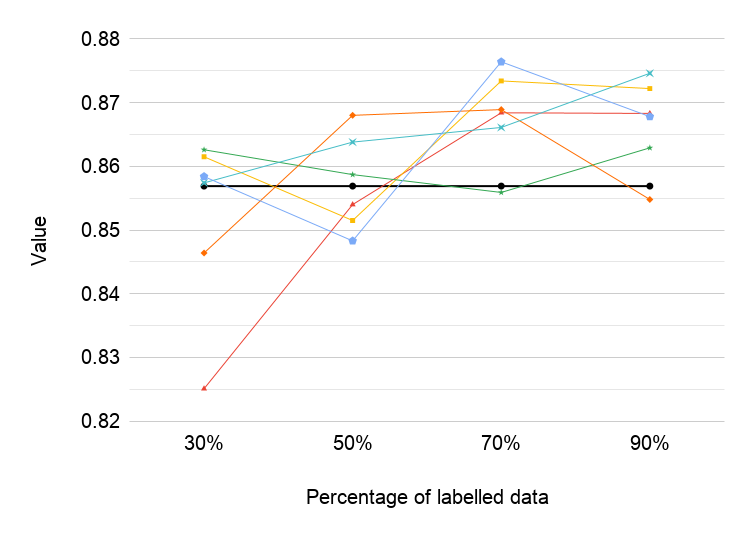}}
	\subfigure[]{\label{auc2}  \includegraphics[width=0.45\textwidth]{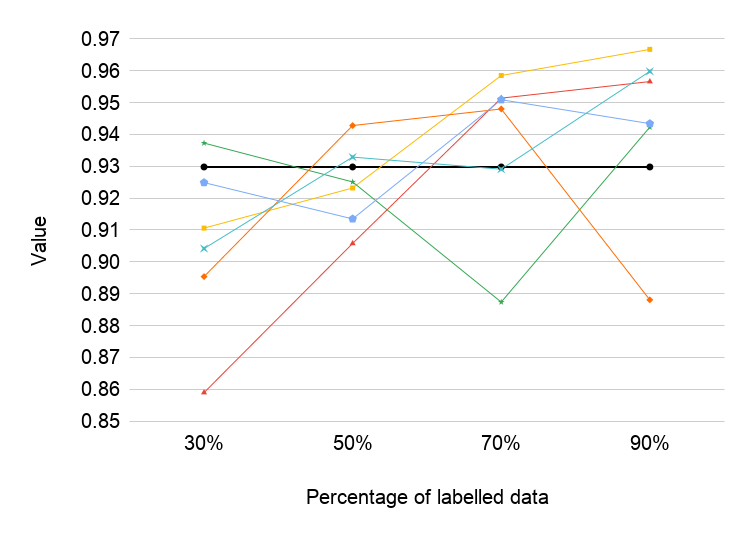}}
	\caption{AUC ROC of the \subref{auc1} internal and \subref{auc2} external test set for the various models.}\label{auc}
\end{center}
\end{figure*}

\begin{table*}
\begin{center}
\begin{tabular}{|l|l|l|l|l|}
\hline
Test Set     & \multicolumn{2}{l|}{Internal Test set} & \multicolumn{2}{l|}{External Test Set}                 \\ \hline
Augmentation & 30\%                       & 50\%                            & 30\%                     & 50\%                                                   \\ \hline
None         & 7.66$\times10^{-12}$ / 1     & 0.4826 / 0.7587         &    2.20 $\times 10^{-16}$ / 1   & 6.04 $\times 10^{-8}$ / 1                              \\ \hline
Translate    &\textbf{0.2288} / 0.1144                  &\textbf{0.176} / 0.912          & 2.34 $\times 10^{-5}$ / 1              & \textbf{0.084} / 0.958       \\ \hline
Rotate       &\textbf{0.1576} / 0.0788       &\textbf{0.6309} / 0.3154            & \textbf{0.0599} / \textbf{0.0300}             & \textbf{0.2606} / 0.8697    \\ \hline
Gamma        & 0.0147 / 0.9927      & 1.86 / \textbf{9.29}$\mathbf{\times10^{-4}}$& 1.11 $\times 10^{-13}$ / 1   & 5.15 $\times 10^{-4}$ / \textbf{2.58} $\mathbf{\times 10^{-4}}$ \\ \hline
Mixup        &\textbf{0.908} / 0.454                 &\textbf{0.0546 / 0.0273}           & 2.70 $\times 10^{-8}$ / 1           & \textbf{0.4118} / 0.2059   \\ \hline
Flip         &\textbf{0.7151} / 0.3575            & 0.02505 / 0.9875               & \textbf{0.2316} / 0.8842               & 6.984 $\times 10^{-4}$ / 0.9997          \\ \hline
\end{tabular}
\end{center}
\caption{p-values of DeLong Test compared to the baseline for the external test set. The first value corresponds to two-tailed DeLong test and the second corresponds to the one-tailed DeLong test. Bold indicates that the value satisfied our criteria, as explained in \nameref{Results}}
\label{tab: pbaseline}
\end{table*}

%\section{Results}
\section{Results}\label{Results}
For comparison of the models' segmentation performance, we plotted each model's mean IoU in Fig. \ref{iou}. We also compared the different models' classification performances by using AUC, in Fig. \ref{auc}.

It is essential to note the behaviour of the NoAug models. In both Fig. \ref{iou} and Fig. \ref{auc}, we see that the performance of NoAug models is lower than baseline for 30\% and 50\% data. However, the NoAug models with 70\% and 90\% data perform as good as, or even better, than the baseline. Thus it is irrelevant to study augmentations on 70\% or more fraction of the data for this study. The p-values for the one-tailed DeLong Test between AUC of NoAug and baseline are $\sim10^{-3}$ and $\sim10^{-4}$ respectively for the internal test set.

To check the first criterion, we did two DeLong hypothesis tests to see if the augmentation models' performance is comparable to the baseline. The first has the null hypothesis that the AUC of augmentation models is equal to that of the baseline trained on 100\% data. In this case, a p-value larger than 0.05 would mean that we fail to reject the null hypothesis, and thus we can say the two AUC are comparable. In the second test, the null hypothesis is that the AUC of the augmentation models is lesser than the baseline. Here a p-value of less than 0.05 would mean that we can reject the null hypothesis, and the AUC of augmentation models is significantly larger than that of the baseline. We find that except for gamma with 30\% data and flip with 50\% data, all models trained on 30\% and 50\% data pass either of the two tests on the internal test set. However, for the external test set, only rotate and flip with 30\% data, and translate, rotate, gamma and mixup with 50\% data pass the hypothesis tests (see table \ref{tab: pbaseline} for the p-values).

We checked the second criterion by performing a one-tailed DeLong Test on the AUC of the augmentation against NoAug models trained on the same data on both test sets. We find that for the external test set, all augmentations with 30\% and 50\% data perform significantly better than NoAug (p $<10^{-2}$). For the internal test set, all augmentations trained on 30\% data performed better than NoAug with 30\% data. However, on 50\% data, only the gamma and mixup performed better than NoAug.

We propose that good augmentations should satisfy both the criteria above for both the test sets. We found that rotate and flip with 30\% data and gamma and mixup with 50\% data accomplished this.

On the other hand, the mean IoU plot informs us of the best data augmentations at pixel-level. As we can see in the internal test set (Fig. \ref{iou1}), rotate and mixup perform quite good at 30\%, whereas mixup and flip perform better at 50\%. For the external test set(Fig. \ref{iou2}), we see that rotate and mixup at 30\% and flip and gamma at 50\% performing better than NoAug and almost as good as the baseline.

Therefore, we find that rotate and mixup are the best augmentations for semantic segmentation on our dataset. These augmentations are capable of reducing the labelled data requirements by even 70\%.

\begin{table}\label{auctable}
\begin{tabular}{|l|c|c|c|c|}
\hline
Test Set          & \multicolumn{2}{l|}{Internal Test set} & \multicolumn{2}{l|}{External Test Set}      \\ \hline
Augmentation      & 30\%               & 50\%              & 30\%                 & 50\%                 \\ \hline
Baseline on 100\% & 0.8569             & 0.8569            & 0.9298               & 0.9298               \\ \hline
None              & 0.8251             & 0.854             & 0.859                & 0.9058               \\ \hline
Translate         & 0.8615             & 0.8515            & 0.9106               & 0.9232               \\ \hline
Rotate            & 0.8626             & 0.8587            & 0.9373               & 0.9251               \\ \hline
Gamma             & 0.8464             & 0.868             & 0.8954               & 0.9428               \\ \hline
Mixup             & 0.8574             & 0.8638            & 0.9042               & 0.9329               \\ \hline
Flip              & 0.8584             & 0.8483            & 0.9249               & 0.9135               \\ \hline
\end{tabular}
\caption{The AUC ROC of the different augmentations calculated on 30\% and 50\% data for both internal and external test sets.}
\end{table}

\section{Conclusion}
Users of deep learning algorithms often overlook image augmentations as an extra step for a minor boost in performance. Our study has shown that augmentations are capable of reducing the amount of labelled training data required. To the best of our knowledge, this is the first study addressing augmentations in this manner.

There is an assumption that augmentations are most useful when the training and test data are from the same distribution \cite{Shorten2019}. Using two test sets from vastly different populations (internal test set being from the RSNA data set, sourced from the USA and the out of sample test set sourced from India) and achieving good results on both, we have shown that this assumption need not hold.

While this study looked at individual augmentations, there is still scope for improving the model performance and reducing the labelled data requirement further by combining multiple augmentations.

% \printbibliography 
\bibliographystyle{ieeetr}
\bibliography{References}
\end{document}